\documentclass[runningheads]{llncs}
\usepackage{graphicx}
\usepackage{booktabs}
\usepackage{amssymb}
\usepackage[space]{cite} 

\usepackage[dvipsnames]{xcolor}

\begin{document}
%
\title{Order of Control and Perceived Control over Personal Information}
\titlerunning{Order of Control and Perceived Control over Personal Information}
%
\author{Yefim Shulman\inst{1}\orcidID{0000-0002-3163-9726}\and
Thao Ngo\inst{2}
\and Joachim Meyer\inst{1}\orcidID{0000-0002-1801-9987}}
\authorrunning{Y. Shulman, T. Ngo and J. Meyer}
%
\institute{Tel Aviv University, Tel Aviv, Israel \\ \email{\{efimshulman@mail.,jmeyer@\}tau.ac.il} \and
University of Duisburg-Essen, Duisburg, Germany \\
\email{thao.ngo@uni-due.de}}

\maketitle              
\begin{abstract}
Focusing on personal information disclosure, we apply control theory and the notion of the \textit{Order of Control} to study people's understanding of the implications of information disclosure and their tendency to consent to disclosure. We analyzed the relevant literature and conducted a preliminary online study (\textit{N} = 220) to explore the relationship between the Order of Control and perceived control over personal information. Our analysis of existing research suggests that the notion of the Order of Control can help us understand people's decisions regarding the control over their personal information. We discuss limitations and future directions for research regarding the application of the idea of the Order of Control to online privacy.

\keywords{Personal Information Disclosure \and Perceived Information Control \and Order of Control \and Privacy 
}
\end{abstract}

\section{Introduction}\label{section:intro}
The desire to control the important aspects of our lives is common, and it may be rooted in nature \cite{kunkel2019integrated}. In this paper, we address one aspect that is becoming increasingly important, namely online privacy regarding the disclosure of personal information.

People's ability to decide which information concerning themselves others should have, and which they should not, is key to the major approach in privacy research that defines ``privacy'' as control over information (e.g.,~\cite{westin1967privacy,warren1890right,parent1983pml,moore2008defining}). A major challenge in this approach is the question whether and how control can be achieved. This is a crucial point in the critique raised by alternative approaches to privacy, such as the theory of ``contextual integrity'', originating in Nissenbaum (2004)~\cite{Nissenbaum2004}.

``Privacy as control'' arises in jurisprudence and ethical philosophy, and, with the technological developments, is more relevant than ever. To overcome the issues related to controllability of information, new tools defining and operationalizing control have been developed, making their way into the legislation. For instance, the traditional triad of information security goals (confidentiality, integrity and availability) was expanded to include three additional goals, specific to privacy protection: unlinkability, transparency and intervenability \cite{hansen2015goals}. This privacy protection triad was intended to address the needs of individuals and society at large. It was originally proposed to form the paradigm of ``linkage control'' \cite{hansen2012topten}, where each of the goals describes an information control relationship between the people (users, data subjects) and data controllers (entities deciding on the purposes and means of personal data processing \cite{eu:gdpr}). Particularly relevant to personal information control are transparency and intervenability. The former states the need for clear descriptions of the intended processing, so people can understand what will be done with the data. The latter goal calls for the ability to intervene in the data processing to allow erasure of data, revision of processing, etc., and to physically exercise control over what may be, can be, and is happening with the data. These two goals are relevant for both the data subjects and the data controllers/processors. Privacy protection through control over personal information is now implemented in legislation, such as the EU's General Data Protection Regulation \cite{eu:gdpr}, which is also applied extraterritorially\footnote{According to the American Bar Association: http://tiny.cc/sf74gz}. 

One of the cornerstones of the discussions of the personal information control in law is the data subject's ability to give and revoke their consent to the data processing. A person's decision to grant consent to access or distribute certain information can be considered a control action, performed by the person to achieve a desired level of exposure of the information. These decisions are informed by the data subjects' perceptions regarding the possible implications from revealing this information. In particular, a person needs to consider whether having knowledge of some information  (perhaps in combination with additional data) allows others to infer other information that the person did not explicitly reveal. Users' perceptions of control are a major determinant of their behavior. Perceived control is linked to privacy concerns (e.g.,~\cite{Malhotra:2004:IUIPC}), and it can elicit risk-compensation in privacy-related scenarios~\cite{Krol:2016:CVE,brandimarte2013controlparadox,Aimeur2016PrivacyPolicyTrust,KOWALEWSKI2015}.


Control theory deals with the formal study of phenomena related to the control of systems and processes. In the previous article, we argued that the framework of control theory can be useful for the analysis of personal information control \cite{Shulman2019Control}. The closed-loop (feedback) control model can serve to describe and achieve privacy protection goals. Control actions, informed by the feedback mechanism, would represent the data processing interventions, informed through ex-ante and ex-post transparency. The informative feedback can provide ex-post transparency to the data subjects, while predictive feedback (or somewhat more peculiar, a feedforward loop) can deliver ex-ante transparency (in line with the discussions on transparency-enhancing technologies \cite{murmann2017}). In our previous work (Shulman and Meyer~\cite{Shulman2019Control}), we presented the conceptual control theoretic analysis of privacy, expressed through personal information disclosure. The conceptual controlled system included:
(1) a person performing actions; (2) a process depending, in part, on these actions; (3) disclosure of personal information as a controlled output; and (4) an evaluation of disclosure. Our analysis emphasized the importance of the control properties of the systems and processes, and their effects on user behavior. One of the relevant properties is the \textit{Order of Control}. This paper presents an analysis of phenomena related to the Order of Control, aiming to assess how it might affect the users' perceptions of control over the disclosure of personal information.

\section{Order of Control for Personal Information}\label{section:orderofcontrol}

\begin{figure}
		\centering
		\includegraphics[width=\textwidth]{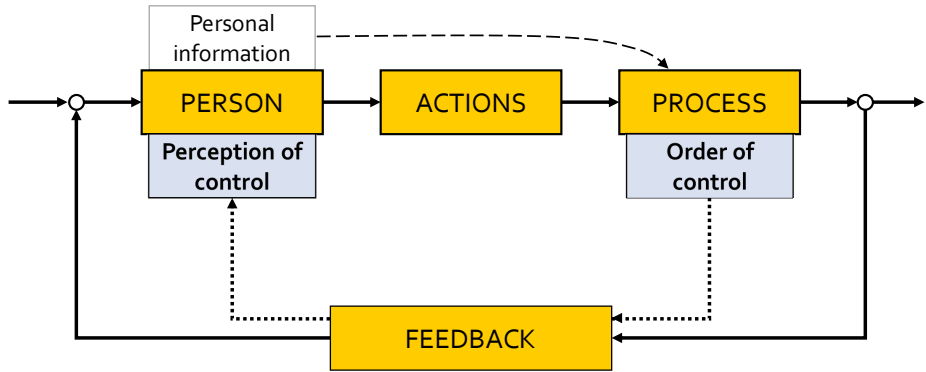}
		\caption{Order of Control as a characteristic of the process affecting the individual's perception of control} \label{fig1}
	\end{figure}

In this work, we rely on, and further develop the ideas first presented in Shulman and Meyer \cite{Shulman2019Control}. We focus on the person (i.e., the data subject), interacting with a process (e.g., using a mobile app, browsing for information, setting up a device, etc.) through some control actions (e.g., pressing buttons) and receiving some feedback on the outcomes of the process (Fig.~\ref{fig1}). The process holder can be another individual or a legal entity who is responsible for the development and/or support of the process or may have administrative access to the process. Personal information is transferred from the person to the process during the interaction. Subsequently, this personal information may be known to the process holder, who assumes the role of the data controller/processor. Ideally, people should have control over their personal information, up to the point where they should be able to have information erased, should they so desire. The GDPR, for instance, mandates the ability of a data subject to give and revoke their consent and fulfil the deletion requests, partly or fully, regarding their personal information \cite{eu:gdpr}. 

In our model, the acts of divulging and deleting personal information constitute control of disclosure, relating to the mechanisms of granular consent, specified by the data protection legislation (i.e., the GDPR). The normative properties of the process (interacting with technology) may affect the person's attitudes and behavior. Specifically, the Order of Control may affect the individual's perception of control, whereby the information about the order itself can be part of the informative feedback the person receives. The Order of Control may serve as a measure of transparency and as a description of the intervenability for the data subjects, affecting the individuals' perception of their ability or agency over their personal information. Additionally, the Order of Control may be applied to analyze privacy protection risks from the standpoint of data controllers or regulators, but in this paper we focus on the individuals', i.e. data subjects', point of view. 


In control theory, the Order of Control is usually defined as the highest order of a differential equation that describes a controlled element. Thus, the Order of Control refers to a dynamic relation between the behavior of the controlled element and the control action, and describes how the input generates the output throughout the control system. This relation may be linear or non-linear, depending on whether the output can be expressed as a linear transformation of the input. The higher the Order of Control, the more difficult it is to control a process, and the more skill is required from the human controller. Thus, the Order of Control of a system or a process defines how much learning will be required from an individual to control it efficiently.

The Order of Control is a fundamental property of the control of a process. We assume that people's own perception of their control over the process should be affected by it. Clearly, the meaning of the Order of Control in the context of privacy differs from the meaning in the context of actual dynamic systems. As we operationalize privacy as personal information disclosure, the most prominent difference is that the Order of Control is a property of the information being disclosed or revealed, together with the processing of this information. We translate the concept of the Order of Control to the domain of personal information disclosure (Table~\ref{tab:orders}) and illustrate it with examples below.

\begin{table}
	\centering
	\caption{Order of Control for personal information disclosure.}\label{tab:orders}
	\begin{tabular}{cll}
	\toprule
		\textbf{Order of} & \textbf{Example from} & \textbf{Example for Personal} \\
			\textbf{Control} & \textbf{Manual Control} & \textbf{Information Disclosure} \\
	\midrule
		0 & Position (displacement). & Disclosing a meaningful and complete \\
			{} & Control of position & item of personal information -- a fact. \\
	\midrule
		1 & Position over time. & Disclosing \{pieces of personal information, \\
			{} & Control of speed -- & accumulation of which over time, \\
				{} & 1\textsuperscript{st} derivative of position & may allow to learn\} the fact. \\
	\midrule
		2 & Speed over time. & Divulging \textbf{[}data, aggregation of which over time, \\
			{} & Control of acceleration -- & may allow to infer \{the pieces of personal \\
				{} & 2\textsuperscript{nd} derivative of position & information, accumulation of which over time, \\
					{} & {} & may allow to learn\}\textbf{]} the fact. \\
	\midrule
		3+ & Jerk, jounce, ``crackle & Divulging $\langle$higher granularity data, collection\\
			{} &  and pop'', etc.& of which over time, may allow to construct \\
				{} & {} & \textbf{[}data, aggregation of which over time, \\
					{} & {} & may allow to infer \{pieces of personal \\
						{}&{}& information, accumulation of which over time, \\
							{}&{}& may allow to learn\}\textbf{]}$\rangle$ the fact \\
	\bottomrule
	\end{tabular}
\end{table}

For a system that requests users' data to deliver its functions: 
\begin{itemize}
	\item Order 0 corresponds to a situation, when a user is asked to disclose a meaningful and complete piece of their personal information, an item (e.g., sexual orientation, some health fact, etc.).
	\item Order 1 corresponds to a request for constituents of personal information (e.g., calendar events), accumulation of which over time may allow to extract a meaningful and complete piece of personal information. For example, from calendar events, such as meetings with other individuals or doctor's appointments, marked in the calendars, a system can draw conclusions about an individual's sexual orientation or health status.
	\item Order 2 corresponds to allowing access to data (e.g., third-party databases, location tracking, etc.) that, when aggregated, may allow to infer a meaningful and complete piece of personal information. For example, a record of locations with timestamps juxtaposed with the data and records of other residents of the area, customers of local businesses, doctors, etc., allows to infer the meetings and appointments, and reveal the sexual orientation and medical diagnoses.
	\item Order 3 and higher may describe, for example, access to the metadata and data of an even higher granularity and lower abstraction level.
\end{itemize}

An important question that follows is how the Order of Control relates to the users' perceived control over their personal information disclosure in interactions with online systems. We conjecture that users' perceptions of control over the process (personal information disclosure) and over their personal information are inversely related to the Order of Control of the controlled process (personal information disclosure) and that of the information. In  other words, the higher the Order of Control, the lower the perceived control over personal information.

\section{Analysis of Literature on Perceptions of Control}\label{section:analysis}

It is possible to analyze the existing literature to gain some insights into the relation between the Order of Control and perceived control. Relying on the methodological recommendations by Webster and Watson \cite{webster2002}, we looked for papers reporting empirical studies\footnote{Note, we refer to ``a paper'' and ``a study'' not as interchangeable terms. Papers study a subject, but report one or more studies. A study is a single complete item of empirical work.} dealing with perceptions of control and personal information disclosure (Table~\ref{tab:selection}). The goal of the literature review was to investigate how the information regarding disclosure and sharing was communicated to the participants in terms of our analysis of the Order of Control of the information. Therefore, we restricted our review to the papers reporting studies, where participants were able to interact with privacy-related interfaces (i.e., privacy policies, permission managers, privacy warnings).
\begin{table}
	\centering
	\caption{Literature survey methodology}\label{tab:selection}
	\begin{tabular}{ll}
		\toprule
		\textbf{Selection criteria}&\textbf{Selection values}\\
		\midrule
			Publication language&English\\
		\midrule
			Venue type&Peer-reviewed journal or conference\\
		\midrule
			Publication period&2004-2019\\
		\midrule
			Publication type&1. Research article (paper) containing research results\\
			&2. Workshop article containing relevant research results\\
		\midrule
			Search queries&1. `perceived control' \& `personal information disclosure'\\
			&2. `perceived control' \& `personal information sharing'\\
		\midrule
			Search query target&Titles, abstracts, full texts (when available)\\
		\midrule
			Search engines&Google Scholar; ACM Digital Library, IEEE Xplore, Scopus\\
			and databases&(see \textit{Search strategy} below)\\
		\midrule
			Initial relevance&Screening of the titles and abstracts\\
		\midrule
			Search \& initial&1. Google Scholar search to acquire relevant initial results\\
			selection strategy&2. Identifying databases containing relevant initial results\\
			&3. Database search to check for further relevant initial results\\
			&4. Lists of references in the initially relevant items\\
			&5. Lists of mentions of the initially relevant items\\
		\midrule
			Inclusion&1. Empirical research on perceived control and personal\\
			requirements&information disclosure; and\\
			&2. Experiment with human subjects as participants; and\\
			&3. Privacy-related user interface (UI) as stimulus material\\
			&(communicating information to the participants); and\\
			&4. The communicated information must contain details related\\
			&to data collection and/or processing; and\\
			&5. Participants ought to interact with the privacy-related UIs\\
		\midrule
			Exclusion bases&Upon further screening: not satisfying inclusion requirements\\
		\bottomrule
	\end{tabular}
\end{table}

The search retrieved 57 initially relevant entries. Overall, after the quality assessment of the search results, we included 16 papers in our analysis (Table~\ref{tab:papers}) that satisfied all the aforementioned criteria (Table~\ref{tab:selection}).

\begin{table}
	\centering
	\caption{Publications included in the analysis.}\label{tab:papers}
	\begin{tabular}{llllll}
		\toprule
		\textbf{Paper} & \textbf{Year} & \textbf{Domain} & \textbf {Method} & \textbf{Control as} & \textbf{Order com-} \\
		{}&{}&{}&{}& \textbf{variable} & \textbf{munication} \\
		\midrule
			Arcand et al.&2007&E-Commerce&2 studies:&Dependent&Privacy\\
			\cite{arcand2007impact}&&&experiments&variable (DV)&policies\\
		\midrule
			Xu&2007&Location-&Experiment&Multiple&On-screen\\
			\cite{xu2007}&&based service&&independent&instructions\\
			&&&&variables (IVs)&\\
		\midrule
			Hoadley et al.&2010&Social&Survey&Self-reported&Facebook feed\\
			\cite{HOADLEY2010}&{}&networks&{}&measure&design change\\
		\midrule
			Lipford et al.&2010&Privacy&Usability&n/a&Privacy\\
			\cite{lipford:2010:VVC}&&settings&study&&policy\\
		\midrule
			Wang et al.&2011&Social&Qualitative&n/a&Permissions\\
			\cite{wang:2011:TAF}&&networks&research&&interface\\
		\midrule
			Brandimarte&2013&Social&3 studies:&IV&Experimental\\
			et al.&&networks&experiments&&description of\\
			\cite{brandimarte2013controlparadox}&&&&&processing/publishing\\
		\midrule
			Christin et al.&2013&Permissions&Usability&IV&Data access\\
			\cite{christin:2013:RUA}&&management&study&&settings\\
		\midrule
			Knijnenburg&2013&Websites,&Experiment&IV&Experimental\\
			et al.&&privacy&&&on-screen\\
			\cite{knijnenburg2013}&&setup&&&information\\
		\midrule
			Keith et al.&2014&Social networks,&Field&IV/DV,&Permission requests,\\
			\cite{keith2014fatigue}&&mobile gaming&study&mediator&privacy settings\\
		\midrule
			Gerlach et al.&2015&Social&Experiment&IV&Privacy\\
			\cite{GERLACH2015}&&networks&&&policy\\
		\midrule
			Tschersich&2015&Social&Experiment&IV&Privacy\\
			\cite{tschersich2015}&&networks&&&settings\\
		\midrule
			Wang et al.&2015&Mobile apps,&Experiment&IV&Permissions\\
			\cite{wang:2015:IEC:2785830.2785845}&&permissions&&&dialogue\\
		\midrule
			A{\"i}meur et al.&2016&E-Commerce&Experiment&IV/DV,&Privacy\\
			\cite{Aimeur2016PrivacyPolicyTrust}&&(allegedly)&&mediator&policies\\
		\midrule
			Steinfeld et al.&2016&Privacy&Experiment&n/a&Privacy policy\\
			\cite{STEINFELD2016}&&policies&&&interaction\\
		\midrule
			Zhang \& Xu&2016&Permissions&Experiment&IV/DV,&Experimental\\
			\cite{zhang:2016:PNM:2818048.2820073}&&management&&mediator&description of\\
			&&&&&access to data\\
		\midrule
			Tsai et al.&2017&Permissions&Usability&n/a&Experimental\\
			\cite{Tsai2017TG}&&management&study&&interface\\
		\bottomrule
	\end{tabular}
\end{table}

The reviewed papers cover various application domains: social networks, permission management in mobile apps, location-based services, privacy policies and privacy settings. Four papers report usability studies of specific tools or designs: \cite{Tsai2017TG,christin:2013:RUA,lipford:2010:VVC,wang:2011:TAF}, while the rest investigate people's attitudes and behaviors in a wider sense: \cite{brandimarte2013controlparadox,HOADLEY2010,GERLACH2015,arcand2007impact,keith2014fatigue,zhang:2016:PNM:2818048.2820073,wang:2015:IEC:2785830.2785845,knijnenburg2013,tschersich2015,Aimeur2016PrivacyPolicyTrust,STEINFELD2016,xu2007}. 

We identified the levels of the Order of Control used in the studies (see Chapter~\ref{section:orderofcontrol}), based on the stimulus material and the design descriptions provided in the papers. Participants could only receive information regarding the Order of Control in their interaction with the stimulus materials and experimental platforms. All studies communicated disclosure-related information at either Order 0: \cite{arcand2007impact,xu2007,lipford:2010:VVC,brandimarte2013controlparadox,knijnenburg2013,GERLACH2015}, or at an Order higher than 2: \cite{arcand2007impact,xu2007,HOADLEY2010,wang:2011:TAF,christin:2013:RUA,keith2014fatigue,GERLACH2015,tschersich2015,zhang:2016:PNM:2818048.2820073,wang:2015:IEC:2785830.2785845,Aimeur2016PrivacyPolicyTrust,STEINFELD2016,Tsai2017TG}, with no studies having intermediate values \footnote{It must be noted that in Xu (2007) \cite{xu2007} the stimuli were designed ad hoc and the description is limited. Therefore, it is not clear whether it corresponds to Order 0 (due to the ``ad hoc'' nature of the experiment) or 1 (as it likely would be in real life circumstances).}. The studies that aimed for ecological validity presented the information similarly to its appearance in the ``real world'', tending to present information of the higher Order. The papers that studied particular factors in models of disclosure presented the information as simple and meaningful as possible, therefore naturally presenting information of the lower Order.

Drawing more general conclusions from the literature proved to be problematic, due to incomparable research questions and hypotheses, differences in the research designs and methods, and unreported effect sizes. Another complication arises, because ``control'' in different papers was used as a dependent or independent variable, moderator or mediator, or as a combination thereof. We can highlight some particular issues in the literature in more detail.

Hoadley et al. \cite{HOADLEY2010} studied people's reaction to Facebook changing the design of its user pages. As Facebook introduced a news feed, which was solely a design change, the Order of Control of information disclosed on the social network did not change. However, the users may have perceived that the Order of Control increased (as Facebook, effectively, started to take the sharing decisions upon itself), and that after the change in the interface design, they had less control over their personal information. This discrepancy in perception has been attributed simply to the way the information regarding disclosure was communicated to the users. The ``Order-of-Control'' interpretation of this observation supports our initial conjecture on the relation of the Order and perception of control (Section~\ref{section:orderofcontrol}).

Gerlach et al. \cite{GERLACH2015} studied how a social network's data-handling practices (as reflected in their privacy policy) relate to the users' willingness to disclose personal information. In a sense, the paper reports a comparison between the information of Order 0 and Orders higher than 2. The authors tested 8 permutations of simplified privacy policy designs, where one of the permutations communicated information with Order 0 and the rest -- with Orders higher than 2. For our purposes, there is a clear asymmetry in the experimental design. However, the results indicate that the effect of the Order of Control might be mediated by privacy risks perceptions, lowering the tendency to disclose personal information.

Brandimarte et al. \cite{brandimarte2013controlparadox} looked into the relations between actual and perceived control, disclosure, and privacy concerns. Their experiments communicated 0 Order of Control to the participants. Their results indicated that when people were unable to estimate the risks of access to, and usage of their personal information (or perceived the risks to be high), and the actual control was perceived to be low, the tendency to disclose decreased. Overall, the higher the perceived control within the same Order of Control, the higher the tendency to disclose.

Arcand et al. \cite{arcand2007impact} investigated the impact of reading privacy policies on the perceptions of control over privacy and trust towards an e-commerce website (through standardized inventories). From our standpoint, the authors tested policies with Order of Control 0 and Order 2 or higher. The policy with ``opt-in'' options represented the lower Order of Control. Compared to the higher Order ``opt-out'' policy, the former one increased perceived control over personal information, supporting our aforementioned conjecture (Section~\ref{section:orderofcontrol}). 

Studying the effects of privacy control complexity on consumer self-disclosure behavior, Keith et al. \cite{keith2014fatigue} argued that perceived ease-of-use determined actual control usage, noting that ``...the more usable the controls, the more likely participants were to adjust the privacy control settings `downward' (make them more restrictive) from the default setting that allowed sharing with all players''. Since the Order of Control in their experiment was higher than 2, the participants might have felt less control, hence trying to restrict personal information disclosure with available actual control. Additionally, Zhang and Xu \cite{zhang:2016:PNM:2818048.2820073} showed that feedback on how permissions were used (with Order higher than 2) might decrease perceived control over personal information.

Overall, the reviewed body of research indicates that our conjecture regarding the relation between the Order of Control and perceptions of control (Section~\ref{section:orderofcontrol}) is justified. One challenge in postulating an effect of the Order of Control would be to show that the control over personal information has levels between what we described as Order 0 and Orders higher than 2. To the best of our knowledge, no systematic study so far looked at the Order of Control in empirical privacy research.


\section{Preliminary Study on the Order of Control and Perceptions of Control}\label{section:orderandperceptions}
 
We conducted a first empirical study to investigate whether there is a relationship between the Order of Control, as a property of information, and the user's perception of control over their personal information. We hypothesized that perceived user control over personal information is inversely related to the Order of Control of the personal information disclosure. In the study, we specifically examine Orders 0, 1 and 2, omitting Orders 3 and higher. This allowed us to simplify the experiment.

In the instructions, we described a fictional home sharing website ``The Platform''. It matches travellers with hosts who are offering a place to stay at their homes. We told the participants that to register for the service, they would need to provide some personal information. Three experimental conditions, representing Orders of Control 0, 1 and 2, differed in the descriptions of the ways in which the information people provide would be processed. In Order 0, the information would be treated as is, without further processing. In Order 1, the information was presented as something that would be accumulated at the request of other users and shared through the Platform. No suggestions were given on what the other users may have wanted to do with the user's personal information. In Order 2, the site stated explicitly that the information could be combined with external or third party databases. We predicted that the participants would consider the possibility that the data they could provide may be used to infer additional information when assessing the sensitivity of providing different types of data. The participants were randomly assigned to one of the three levels of the Order of Control.


\subsection{Method}
\subsubsection{Participants}
All in all, 220 participants (age range: 18 - 69, 39.1\% female, 0.9\% undisclosed gender) were recruited through the crowd-sourcing platform Amazon Mechanical Turk and completed the task. The self-reported educational levels varied from no degree to Master's and professional degrees, with the majority holding a Bachelor's degree or its equivalent. The distribution of the participants between the three Order of Control groups was comparable (75, 76 and 69 people in each group, respectively).

\subsubsection{Study Design.}
We applied a between-subject design, using the \textit{Order of Control} as an independent variable with three conditions related to the different Orders of Control described in
Table~\ref{tab:orders}: 
\begin{enumerate}
\item Order of Control 0, referring to personal information disclosure as revealing of a fact;
\item Order of Control 1, referring to disclosure of the elements of personal information over time -- constituting elements of a new fact; 
\item Order of Control 2, referring to disclosure of the data over time, combination of which with other data constitutes the elements, from which a new fact can be inferred.
\end{enumerate}

As a dependent variable, we measured perceived controllability over personal information, perceived information control, as well as online and physical privacy concerns through standardized questionnaires. 

\subsubsection{Materials.}\label{section:materials} 
Our experiment contained several measures, some of which were based on tools we adapted from the literature.

\textbf{Perceived controllability.} To measure the perceived controllability of personal information, we created a list of 15 personal information items. This list was also used in the scenarios, presented at the beginning of the experiment. After the presentation of the scenarios, the participants had to respond to the question of ``How easy will it be for you to control the access and use of the personal information that you may disclose through the platform?'' on a 9-point Likert-type scale including the ``\textit{I do not want to answer}'' option. 
The 15 items included: 
\begin{itemize}
\item First name;
\item Last name;
\item Email address;
\item Phone number;
\item A personal photo;
\item National ID, residence permit or equivalent;
\item Address;
\item Links to social media accounts (e.g., Facebook, Twitter);
\item Hobbies;
\item Countries visited so far;
\item Photos of the apartment;
\item The amount of rent paid per month;
\item Sleeping time;
\item The time at home;
\item Favorite places in the city.
\end{itemize}

\textbf{Perceived information control.}
Four items were used to measure perceived control over personal information, adapted from Dinev et al.~\cite{DinevXu2013IPC}, e.g., ``I think I have control over what personal information is released by the home sharing platform.'' The items were measured on a 7-point Likert-type scale, ranging from ``1 = I do not agree at all'' to ``7 = I fully agree''. The internal consistency of this measure in the original paper was reported as good (Cronbach's alpha = .89).

\textbf{Online privacy concerns.}
Four items were used to measure online privacy concerns, adapted from Lutz et al.~\cite{Lutz2017RolePrivacy}. The participants were asked to indicate their level of concern about potential online privacy risks that could arise from personal information disclosure on the home sharing platform (the risk of identity theft, hacking, cyberstalking and publishing personal information without consent). The items were measured on a 5-point Likert-type scale, ranging from ``1 = No concern at all'' to ``5 = Very high concern''. The internal consistency of the measurement in the original paper was reported as good (Cronbach's alpha = .92).

\textbf{Physical privacy concerns.}
To measure physical privacy concerns, we used four items adapted from Lutz et al.~\cite{Lutz2017RolePrivacy}. The participants had to indicate their level of concern about potential privacy risks that could arise when hosting somebody via a home sharing platform. These risks included damage to personal belongings, the guest snooping through personal belonging, the guest entering personal areas without permission, and the guest using items that should not be used. The items ranged from ``1 = No concern at all'' to ``5 = Very high concern''. The internal consistency of the measurement in the original paper was reported as good (Cronbach's alpha = .89).

\subsection{Procedure}
The participants were randomly assigned to one of the three between-subject conditions, corresponding to the three Orders of Control (Table~\ref{tab:orders}). In each condition, a scenario about a registration on a novel (fictive) home sharing platform was presented in text. The scenarios were introduced as follows:

\textit{Imagine you are about to register on and start using a home sharing website, called ``The Platform''. You can use it, when travelling yourself, to stay at other users' homes, or you can host other travellers in your own home, or both. If you decide to use the service, a profile will be created for you on The Platform. Your profile will only be accessible by you and by the members of this respective home sharing community, who registered the same way as you did. Your profile information will be disclosed to the home sharing platform itself. In order to use the home sharing platform, some information has to be shared with other users of The Platform.}

\begin{figure}
        \center{\includegraphics[width=\textwidth]{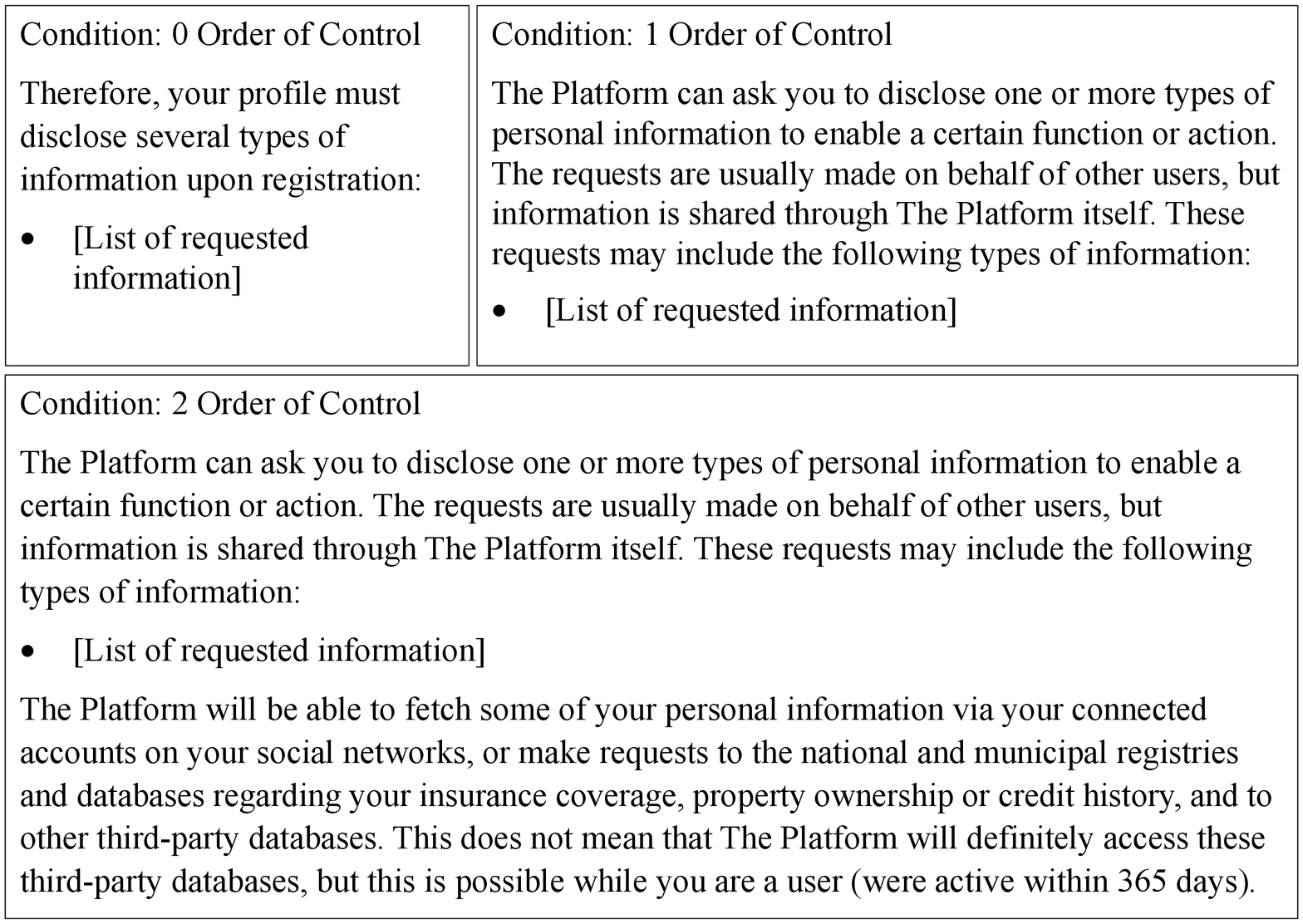}}
        \caption{Manipulation of the Order of Control for each scenario}
        \label{fig2}
        
    \end{figure}

The list of requested information consisted of the 15 personal information items. We manipulated the Order of Control through the description of the way the information would be used (Figure~\ref{fig2}). 

The experiment flow was as follows:
\begin{enumerate}
    \item Greetings and informed consent form.
    \item Presentation of a scenario per randomly assigned condition.
    \item Items measuring the dependent variable.
    \item Measurement of information control perceptions and privacy concerns.
    \item General demographics questions and attention check.
    \item Debriefing. The participants received monetary compensation for participation through Amazon Mechanical Turk. 
\end{enumerate}

\subsection{Results}
We performed an exploratory factor analysis (EFA) on the 15 items measuring perceived information control. The EFA (with primary axis factoring and oblimin rotation) revealed that 13 of our 15 personal information items reliably reflected 2 factors (Cronbach's alpha of .93 and .90, respectively): \textit{``control over disclosing identity''} (first and last name, email, phone, address, personal photo, national ID, and photos of the apartment); and \textit{``control over disclosing preferences''} (sleeping schedule, favorite places in the city, visited countries, home presence time, and hobbies). The KMO measure was not significant $>.8$, and Bartlett's test of sphericity was significant $<.001$. The links to social media and the amount of monthly rent did not load sufficiently and were excluded from further analyses.   

We analyzed the relation of the \textit{Order of Control} and \textit{perceived information control} with several configurations of analysis of variance (ANOVA, MANOVA, mixed ANOVA, testing our hypothesis from different angles), finding no significant effect of the former at $p<.05$. 

\subsection{Discussion}\label{section:discussion}
In our preliminary study, we aimed to investigate the relationship between the Order of Control and perceived control over personal information. The results revealed no significant effect of the Order of Control on the perceptions of control over personal information. However, the exploratory factor analysis, which we performed to explore latent variables, revealed that perceived control over personal information items can be meaningfully split into two categories, which we termed: \textit{control over disclosing identity} and \textit{control over disclosing preferences}. This finding might be helpful for conceptualizing Order of Control in the context of privacy. For instance, we can now hypothesize that \textit{control over disclosing identity} might differ in execution and perception from \textit{control over disclosing preferences}, as the latter occurs as unintended information disclosure (e.g., online search, or clicking behavior). Therefore, the relation between the Order of Control and perceptions of control may differ for different types of personal information. This may be a topic for future research. 

In our experiment, we found no significant effect of the Order of Control on perceived control over personal information. This null-effect could be due to one of two causes. First, the Order of Control may have little value in the context of privacy and control over personal information. Second, our study had methodological shortcomings that caused our results to be non-significant. In particular, our manipulation of the Order of Control may have been inadequate. We did not conduct a systematic manipulation check, so we do not know if the participants indeed perceived the different conditions as different Orders of Control. Future research needs to address this shortcoming.

Specifically, in future studies, qualitative pre-tests could be run before the main study to ensure the comprehension and readability of the scenarios. The manipulation check should also help to understand the potential outcomes of the experiment. The conceptual conjecture regarding the Order of Control may be split into separate hypotheses to explicitly test its underlying components. At a technical level, the difference between the Orders of Control in the experiment may not have been clear enough to obtain a significant effect. 

\section{General Discussion and Conclusion}\label{section:conclusion}

In this paper, we extend our previously proposed approach to study privacy-related phenomena through the lens of the control theoretic framework~\cite{Shulman2019Control}. We added an analysis of the effects of intrinsic properties of controlled systems on users' attitudes towards control. We focus on and conceptualize the Order of Control as one such property of information disclosure in online systems.

We looked at how the Order of Control relates to the users' perceptions of control over their personal information. We conjectured an inverse relation between the Order of Control and the users' perception of control, and described the necessary steps to investigate the proposed relation. The analysis of the relevant literature supported our idea regarding the Order of Control and highlighted the challenges arising from the delicate interplay of actual control, perceived control, risk perception, feedback presentation and personal information disclosure.

We also attempted to study this relation with an online experiment. Our analysis revealed two categories of perceived control: \textit{control over disclosing identity}, and \textit{control over disclosing preferences}. This distinction may have profound implications for the meaningful communication of privacy-related information via notices and indications, for individuals' privacy risk perceptions, and for the efficient application of transparency-enhancing technologies. These potential implications may warrant future research. 

We found no significant effect of the Order of Control on perceptions of control. This, in our opinion, highlights some methodological challenges in the study of the effects of the Order of Control and related issues in the context of privacy. Further research is needed to establish a better understanding of the determinants of the controllability and perceptions of control over personal information.

Finally, having focused on the individuals' perceptions in this paper, we can see another promising direction for future research. The Order of Control can also provide a measure of granularity of the personal information disclosed to the online systems, and, potentially, to the data controllers/processors. It can characterize data processing breadth and depth, serving as the basis for the evaluation of the information inference. As a property of information, the Order of Control can be used to evaluate transparency and to describe how information may affect the individuals' perceptions of intervenability. Furthermore, the Order of Control may help system designers and auditors estimate transparency as a whole and quantify intervenability of data processing for a given information system interaction. Thus, the Order of Control may not only be a useful concept for studying the individuals' attitudes and behaviors, but it may also be relevant for data controllers and regulators, performing privacy risks assessment (``data protection impact assessment'' under the GDPR \cite{bieker2016DPIA}) procedures.

	\subsubsection{Funding.}
This research is partially funded by the EU Horizon 2020 research and innovation programme under the Marie Sk{\l}odowska-Curie grant agreement No 675730 ``Privacy and Us'', and by the Deutsche Forschungsgemeinschaft (DFG) under Grant No. GRK 2167, Research Training Group ``User-Centered Social Media''.

	\bibliographystyle{splncs04}
\bibliography{bibliography_OOC}

\begin{thebibliography}{10}
\providecommand{\url}[1]{\texttt{#1}}
\providecommand{\urlprefix}{URL }
\providecommand{\doi}[1]{https://doi.org/#1}

\bibitem{Aimeur2016PrivacyPolicyTrust}
A{\"i}meur, E., Lawani, O., Dalkir, K.: When changing the look of privacy
  policies affects user trust: An experimental study. Computers in Human
  Behavior  \textbf{58},  368 -- 379 (2016). \doi{10.1016/j.chb.2015.11.014}

\bibitem{arcand2007impact}
Arcand, M., Nantel, J., Arles-Dufour, M., Vincent, A.: The impact of reading a
  web site's privacy statement on perceived control over privacy and perceived
  trust. Online Information Review  \textbf{31}(5),  661--681 (2007)

\bibitem{bieker2016DPIA}
Bieker, F., Friedewald, M., Hansen, M., Obersteller, H., Rost, M.: A process
  for data protection impact assessment under the european general data
  protection regulation. In: Schiffner, S., Serna, J., Ikonomou, D.,
  Rannenberg, K. (eds.) Privacy Technologies and Policy. pp. 21--37. Springer
  International Publishing, Cham (2016)

\bibitem{brandimarte2013controlparadox}
Brandimarte, L., Acquisti, A., Loewenstein, G.: Misplaced confidences: Privacy
  and the control paradox. Social Psychological and Personality Science
  \textbf{4}(3),  340--347 (2013). \doi{10.1177/1948550612455931}

\bibitem{christin:2013:RUA}
Christin, D., Michalak, M., Hollick, M.: Raising user awareness about privacy
  threats in participatory sensing applications through graphical warnings. In:
  Proceedings of International Conference on Advances in Mobile Computing \&
  Multimedia. pp. 445:445--445:454. MoMM '13, ACM, New York, NY, USA (2013).
  \doi{10.1145/2536853.2536861}

\bibitem{DinevXu2013IPC}
Dinev, T., Xu, H., Smith, J.H., Hart, P.: Information privacy and correlates:
  an empirical attempt to bridge and distinguish privacy-related concepts.
  European Journal of Information Systems  \textbf{22}(3),  295--316 (2013).
  \doi{10.1057/ejis.2012.23}

\bibitem{eu:gdpr}
{EU 2016/679}: Regulation (eu) 2016/679 of the european parliament and of the
  council of 27 april 2016 on the protection of natural persons with regard to
  the processing of personal data and on the free movement of such data, and
  repealing directive 95/46/ec (general data protection regulation). Official
  Journal of the European Union  \textbf{L119},  1--88 (May 2016),
  \url{http://eur-lex.europa.eu/legal-content/EN/TXT/?uri=OJ:L:2016:119:TOC}

\bibitem{GERLACH2015}
Gerlach, J., Widjaja, T., Buxmann, P.: Handle with care: How online social
  network providers’ privacy policies impact users’ information sharing
  behavior. The Journal of Strategic Information Systems  \textbf{24}(1),  33
  -- 43 (2015). \doi{10.1016/j.jsis.2014.09.001}

\bibitem{hansen2015goals}
{Hansen}, M., {Jensen}, M., {Rost}, M.: Protection goals for privacy
  engineering. In: 2015 IEEE Security and Privacy Workshops. pp. 159--166 (May
  2015). \doi{10.1109/SPW.2015.13}

\bibitem{hansen2012topten}
Hansen, M.: Top 10 mistakes in system design from a privacy perspective and
  privacy protection goals. In: Camenisch, J., Crispo, B., Fischer-H{\"u}bner,
  S., Leenes, R., Russello, G. (eds.) Privacy and Identity Management for Life.
  pp. 14--31. Springer Berlin Heidelberg, Berlin, Heidelberg (2012)

\bibitem{HOADLEY2010}
Hoadley, C.M., Xu, H., Lee, J.J., Rosson, M.B.: Privacy as information access
  and illusory control: The case of the facebook news feed privacy outcry.
  Electronic Commerce Research and Applications  \textbf{9}(1),  50 -- 60
  (2010). \doi{10.1016/j.elerap.2009.05.001}, special Issue: Social Networks
  and Web 2.0

\bibitem{keith2014fatigue}
Keith, M., Maynes, C., Lowry, P., Babb, J.: Privacy fatigue: The effect of
  privacy control complexity on consumer electronic information disclosure (12
  2014). \doi{10.13140/2.1.3164.6403}

\bibitem{knijnenburg2013}
Knijnenburg, B.P., Kobsa, A., Jin, H.: Counteracting the negative effect of
  form auto-completion on the privacy calculus. In: 34th International
  Conference on Information Systems. Milan, Italy (Dec 15-18 2013)

\bibitem{KOWALEWSKI2015}
Kowalewski, S., Ziefle, M., Ziegeldorf, H., Wehrle, K.: Like us on facebook!
  – analyzing user preferences regarding privacy settings in germany.
  Procedia Manufacturing  \textbf{3},  815 -- 822 (2015).
  \doi{10.1016/j.promfg.2015.07.336},
  \url{http://www.sciencedirect.com/science/article/pii/S2351978915003376}, 6th
  International Conference on Applied Human Factors and Ergonomics (AHFE 2015)
  and the Affiliated Conferences, AHFE 2015

\bibitem{Krol:2016:CVE}
Krol, K., Preibusch, S.: Control versus effort in privacy warnings for
  webforms. In: Proceedings of the 2016 ACM on Workshop on Privacy in the
  Electronic Society. pp. 13--23. WPES '16, ACM, New York, NY, USA (2016).
  \doi{10.1145/2994620.2994640}

\bibitem{kunkel2019integrated}
Kunkel, J., Luo, X., Capaldi, A.P.: Integrated torc1 and pka signaling control
  the temporal activation of glucose-induced gene expression in yeast. Nature
  communications  \textbf{10}(1),  1--11 (2019)

\bibitem{lipford:2010:VVC}
Lipford, H.R., Watson, J., Whitney, M., Froiland, K., Reeder, R.W.: Visual vs.
  compact: A comparison of privacy policy interfaces. In: Proceedings of the
  SIGCHI Conference on Human Factors in Computing Systems. pp. 1111--1114. CHI
  '10, ACM, New York, NY, USA (2010). \doi{10.1145/1753326.1753492}

\bibitem{Lutz2017RolePrivacy}
Lutz, C., Hoffmann, C.P., Bucher, E., Fieseler, C.: The role of privacy
  concerns in the sharing economy. Information, Communication \& Society
  \textbf{21}(10),  1472--1492 (2018). \doi{10.1080/1369118X.2017.1339726}

\bibitem{Malhotra:2004:IUIPC}
Malhotra, N.K., Kim, S.S., Agarwal, J.: Internet users' information privacy
  concerns (iuipc): The construct, the scale, and a causal model. Info. Sys.
  Research  \textbf{15}(4),  336--355 (Dec 2004). \doi{10.1287/isre.1040.0032}

\bibitem{moore2008defining}
Moore, A.: Defining privacy. Journal of Social Philosophy  \textbf{39}(3),
  411--428 (2008)

\bibitem{murmann2017}
{Murmann}, P., {Fischer-H{\"u}bner}, S.: Tools for achieving usable ex post
  transparency: A survey. IEEE Access  \textbf{5},  22965--22991 (2017).
  \doi{10.1109/ACCESS.2017.2765539}

\bibitem{Nissenbaum2004}
Nissenbaum, H.: Privacy as contextual integrity. Washington Law Review
  \textbf{79}(1),  119--157 (2004),
  \url{https://www.scopus.com/inward/record.uri?eid=2-s2.0-1842538795\&partnerID=40\&md5=377fc1b3e8b0a416836505aaea590b01}

\bibitem{parent1983pml}
Parent, W.A.: Privacy, morality, and the law. Philosophy \& Public Affairs
  \textbf{12}(4),  269--288 (1983), \url{http://www.jstor.org/stable/2265374}

\bibitem{Shulman2019Control}
Shulman, Y., Meyer, J.: Is privacy controllable? In: Kosta, E., Pierson, J.,
  Slamanig, D., Fischer-H{\"u}bner, S., Krenn, S. (eds.) Privacy and Identity
  Management. Fairness, Accountability, and Transparency in the Age of Big
  Data: 13th IFIP WG 9.2, 9.6/11.7, 11.6/SIG 9.2.2 International Summer School,
  Vienna, Austria, August 20-24, 2018, Revised Selected Papers. pp. 222--238.
  IFIP Advances in Information and Communication Technology, Springer
  International Publishing, Cham (2019). \doi{10.1007/978-3-030-16744-8\_15},
  the authors version is available in open access via:
  \url{http://arxiv.org/abs/1901.09804}

\bibitem{STEINFELD2016}
Steinfeld, N.: “i agree to the terms and conditions”: (how) do users read
  privacy policies online? an eye-tracking experiment. Computers in Human
  Behavior  \textbf{55},  992 -- 1000 (2016). \doi{10.1016/j.chb.2015.09.038}

\bibitem{Tsai2017TG}
Tsai, L., Wijesekera, P., Reardon, J., Reyes, I., Egelman, S., Wagner, D.,
  Good, N., Chen, J.W.: Turtle guard: Helping android users apply contextual
  privacy preferences. In: Thirteenth Symposium on Usable Privacy and Security
  ({SOUPS} 2017). pp. 145--162. {USENIX} Association, Santa Clara, CA (2017),
  \url{https://www.usenix.org/conference/soups2017/technical-sessions/presentation/tsai}

\bibitem{tschersich2015}
{Tschersich}, M.: Comparing the configuration of privacy settings on social
  network sites based on different default options. In: 2015 48th Hawaii
  International Conference on System Sciences. pp. 3453--3462 (Jan 2015).
  \doi{10.1109/HICSS.2015.416}

\bibitem{wang:2011:TAF}
Wang, N., Xu, H., Grossklags, J.: Third-party apps on facebook: Privacy and the
  illusion of control. In: Proceedings of the 5th ACM Symposium on Computer
  Human Interaction for Management of Information Technology. pp. 4:1--4:10.
  CHIMIT '11, ACM, New York, NY, USA (2011). \doi{10.1145/2076444.2076448}

\bibitem{wang:2015:IEC:2785830.2785845}
Wang, N., Zhang, B., Liu, B., Jin, H.: Investigating effects of control and ads
  awareness on android users' privacy behaviors and perceptions. In:
  Proceedings of the 17th International Conference on Human-Computer
  Interaction with Mobile Devices and Services. pp. 373--382. MobileHCI '15,
  ACM, New York, NY, USA (2015). \doi{10.1145/2785830.2785845}

\bibitem{warren1890right}
Warren, S.D., Brandeis, L.D.: The right to privacy. Harvard law review pp.
  193--220 (1890)

\bibitem{webster2002}
Webster, J., Watson, R.T.: Analyzing the past to prepare for the future:
  Writing a literature review. MIS Quarterly  \textbf{26}(2),  xiii--xxiii
  (2002), \url{http://www.jstor.org/stable/4132319}

\bibitem{westin1967privacy}
Westin, A.: Privacy and freedom. 1967. Atheneum, New York  (1970)

\bibitem{xu2007}
Xu, H.: The effects of self-construal and perceived control on privacy
  concerns. In: ICIS 2007 Proceedings - Twenty-Eighth International Conference
  on Information Systems (2007)

\bibitem{zhang:2016:PNM:2818048.2820073}
Zhang, B., Xu, H.: Privacy nudges for mobile applications: Effects on the
  creepiness emotion and privacy attitudes. In: Proceedings of the 19th ACM
  Conference on Computer-Supported Cooperative Work \& Social Computing. pp.
  1676--1690. CSCW '16, ACM, New York, NY, USA (2016).
  \doi{10.1145/2818048.2820073}

\end{thebibliography}
\end{document}